\documentclass{emulateapj}

\usepackage{epsfig}

\newcommand{\novamon}{A0620--00}

\newcommand{\Halpha}{H$\alpha$}

\newcommand{\Hbeta}{H$\beta$}

\slugcomment{ApJ Submission}

\shorttitle{Disks in Quiescent Black Hole Binaries}
\shortauthors{Hynes et al.}

\begin{document}


\title{Observational Constraints Upon Cool
  Disk Material in Quiescent Black Hole Binaries}

\author{R. I. Hynes\altaffilmark{1,2}, E. L. Robinson\altaffilmark{2}, 
M. Bitner\altaffilmark{2}}
\altaffiltext{1}{Department of Physics and
  Astronomy, Louisiana State University, Baton Rouge,
  Louisiana 70803, USA; rih@phys.lsu.edu}
\altaffiltext{2}{McDonald Observatory and Department of Astronomy, The
University of Texas at Austin, 1 University Station C1400, Austin,
Texas 78712-0259}

\begin{abstract}

We consider current observational constraints on the presence of cool,
optically thick disk material in quiescent black hole binaries,
specifically focusing on a case study of the prototypical system
A\,0620--00.  Such material might be expected to be present
theoretically, but is usually claimed to make a negligible
contribution at optical and infrared wavelengths.  The primary
argument is based on measurements of the veiling of stellar
photospheric absorption lines, in which it is assumed that the disk
spectrum is featureless.  We use simulated spectra to explore the
sensitivity of veiling measurements to uncertainties in companion
temperature, gravity, and metallicity.  We find that the derived
veiling is extremely sensitive to a mismatch between the temperature
and metallicity of the companion and template, but that the effect of
a plausible gravity mismatch is much smaller.  In general the
resulting uncertainty in the amount of veiling is likely to be much
larger than the usually quoted statistical uncertainty.  We also
simulate spectra in which the disk has an emergent spectrum similar to
the star and find that in this case, optical veiling constraints are
moderately robust.  This is because the rotational broadening of the
disk is so large that the two line profiles effectively decouple and
the measurement of the depth of stellar lines is largely unbiased by
the disk component.  We note, however, that this is only true at
intermediate resolutions or higher, and that significant bias might
still affect low resolution IR observations.  Assuming that the
optical veiling is reliable, we then examine the constraints upon the
temperature and covering factor of any optically thick disk component.
These are stringent if the disk is warm ($T_{\rm eff} \ga 3500$\,K),
but very temperature sensitive, and cooler disks are largely
unconstrained by optical measurements.  Current IR veiling estimates
do not help much, representing rather high upper limits.  Probably the
best constraint comes from the relative amplitudes of ellipsoidal
variations in different bands as these are sensitive to differences in
veiling which are expected for disks cooler than the companion star.
A significant disk contribution in the IR, up to $\sim25$\,\%, is not
ruled out in this or any other way considered, however.

\end{abstract}


\section{Introduction}

Almost all of our knowledge of the masses of stellar-mass black holes
derives from those in low mass X-ray binaries: the black hole X-ray
transients (BHXRTs; e.g.\ \citealt{Cherepashchuk:2000a}).  In most of
these systems substantial X-ray activity occurs only during well
defined outbursts.  Between outbursts, the emission from the accretion
flow fades to the point that the companion star is clearly visible and
nearly undisturbed by irradiation; hence it can be used to derive a
dynamical mass for the black hole.  Current methodology relies on
three key ingredients to determine a mass.  Firstly the radial
velocity curve of the companion star yields the binary period (if this
is not already known), and the companion's projected radial velocity
semi-amplitude.  Based on these quantities alone, the mass function
can be calculated, and this sets an absolute minimum mass for the
compact object.  Secondly the rotational broadening of the companion
star's absorption lines is used to derive the mass ratio.  Finally,
using photometric observations (preferably in the IR) of the
ellipsoidal modulation of light from the companion star, coupled with
models of this modulation the binary inclination can be determined as
a function of assumed mass ratio.  Together these quantities yield the
mass of the black hole.

In the ideal case there will be no light at all from the accretion
flow.  This is never completely the case, however, and we usually must
make do with making some assumptions about its effect on the analysis.
Typically it is assumed that the spectrum is a featureless continuum
and that its contribution is minimized in the IR.  These assumptions
are questionable, however.  The disk instability model (DIM;
\citealt{Lasota:2001a}) advanced to explain the transient behavior
predicts that the quiescent disk should be very cool, likely cooler
than the companion star, and could be optically thick
\citep{Cannizzo:1993a,Menou:2002a}.  It has been argued that such a
disk is virtually required to account for the necessary surface
densities, although we argue in Section~\ref{DiscussionSection} that
this requirement is much weaker than usually stated.  Under these
conditions we might expect both a significant IR contribution and
possibly absorption lines similar to those of a late-type star.
Unaccounted for disk emission will bias inclination determinations in
the sense that inclinations will be underestimated if disk light is
not taken into account.  This effect is quantified by measuring the
dilution of the photospheric absorption lines, and is almost always
done in the optical.  One must then ask how robust this estimate of
the dilution (the veiling) is, whether it is biased by absorption
lines from the disk, and whether the veiling from a cool disk could be
significantly larger in the IR than in the optical.

We know that a disk is present in quiescence.  This is indicated by
line profiles and Doppler tomography of emission lines (e.g.\
\citealt{Marsh:1994a}) and probably by observations of flickering
behavior (e.g.\ \citealt{Zurita:2003a}; \citealt{Hynes:2003a}).  The
line observations, at least, are related to hotter and optically thin
gas which probably co-exists with the cool optically thick material
that we are considering, either in a chromospheric layer above the
disk, or in low surface density regions \citep{Vrielmann:2002a}.  The
origin of the flickering is less clear.  In at least one case,
V404~Cyg, it has a line component which appears to be distributed
across the disk and driven by X-ray heating
\citep{Hynes:2002a,Hynes:2004b}.  These flares exhibit a continuum as
well, but this too may be optically thin, and/or associated with the
inner disk.  None of these observations inform us about a possible
cooler optically thick component of the disk.  We also note that in
the quiescent white dwarf system, WZ~Sge, \citet{Howell:2004a} have
recently identified CO and H$_2$ in {\em emission}.  The authors argue
that this originates in a cool (2000--5000\,K) component of the disk.
No equivalent detection has yet been made in quiescent BHXRTs, and
WZ~Sge is a unique object that may not be comparable to them, but this
should be borne in mind.

In this work, we will consider what other evidence exists to constrain
such cool material.  We will focus on one system as a case-study, the
prototypical BHXRT, A\,0620--00; this work is intended to be an
example of the severity of different sources of uncertainty rather
than an exhaustive exploration of parameter space.  We will begin by
reviewing the relevant system parameters.  We then construct synthetic
line profiles, and model spectra, for both the companion star and an
optically thick disk with a cool, stellar spectrum.  Then we use these
to test how sensitive veiling measurements are to uncertainties in the
temperature, gravity, and metallicity of the companion, and whether
the veiling measurements may be biased by the presence of such disk
absorption lines.  Assuming that the optical veiling estimates are
reliable, we next examine the constraints this places upon the
quiescent disk.  Finally we will examine potentially more stringent
constraints from the IR veiling and photometric color.

\section{A\,0620--00: Parameters for a Case Study}

A\,0620--00 is one of the best studied systems with both optical
\citep{Marsh:1994a} and IR \citep{Shahbaz:1999a} veiling estimates as
well as recent high-resolution optical spectroscopy
\citep{Gonzalez:2004a}.

The temperature of the companion star is critical for veiling
estimates, but is typically very uncertain in BHXRTs.  Several of the
methods used may be compromised if the disk contamination is
significant, and does not vary monotonically with wavelength.  In
particular, photometric spectral types (e.g.\ K4V,
\citealt{Gelino:2001a}) could be significantly biased by contamination
of the continuum by another source of light, a point also appreciated
by \citet{Oke:1977a}.  Similarly spectrophotometric constraints (e.g.\
K3V-K4V; \citealt{Haswell:1992a}) have the same problem.  Constraints
based purely upon spectral features are relatively weak for \novamon,
with the exception of the recent work by \citet{Gonzalez:2004a} which
we will discuss later.  Most have used broad bands rather than sharp
features.  \citet{Oke:1977a} obtained K5--K7, and preferred a dwarf to
a giant.  Subsequent observations have found molecular bands to be
weaker, however, and allow earlier types.  \citet{McClintock:1986a}
used K2V, K3V, K4V, and K7V templates and did not explicitly reject
any of them.  \citet{Haswell:1992a} preferred K3V or K4V, but could
not reject K8V.  Earlier spectral types are possible as the only other
type considered (and rejected) was G8V.  \citet{Marsh:1994a} only used
a K3V--K4V template.  Finally \citet{Shahbaz:1999a} examined the CO
bandhead in the $K$ band and compared to K0V, K3V, K5V, and K7V
spectral types.  K3V was preferred, but nearby types could not be
ruled out, especially as \citet{Froning:2001a} challenged the validity
of results based purely on the CO bandhead because of its sensitivity
to surface gravity and metallicity.  In summary, then, most studies
allow a range of spectral types.  A classification as early as K2V, or
perhaps even K1V, cannot be definitely rejected.  Similarly later
types, K4V--K5V, perhaps as high as K7V--K8V are not implausible
considering only spectroscopic evidence.  For this work we will assume
K4V, and a temperature of $T_{\rm eff}=4690$\,K \citep{Gray:1994a}.
We have belabored this point more than strictly necessary to
illustrate the uncertainties present even for a well studied object.

The more recent analysis by \citet{Gonzalez:2004a} is superior to
those discussed above, and yields a somewhat hotter temperature,
$T_{\rm eff}=4900\pm100$\,K, corresponding to a spectral type of K3V.
We deliberately do not use this value in our simulations as it is not
representative of the quality of measurement available in other
systems.  We do keep it in mind as a further indication that typical
temperature estimates are uncertain by at least 200\,K (and often
more) and will comment further where appropriate.


\begin{figure*}
\epsfig{width=2.4in,file=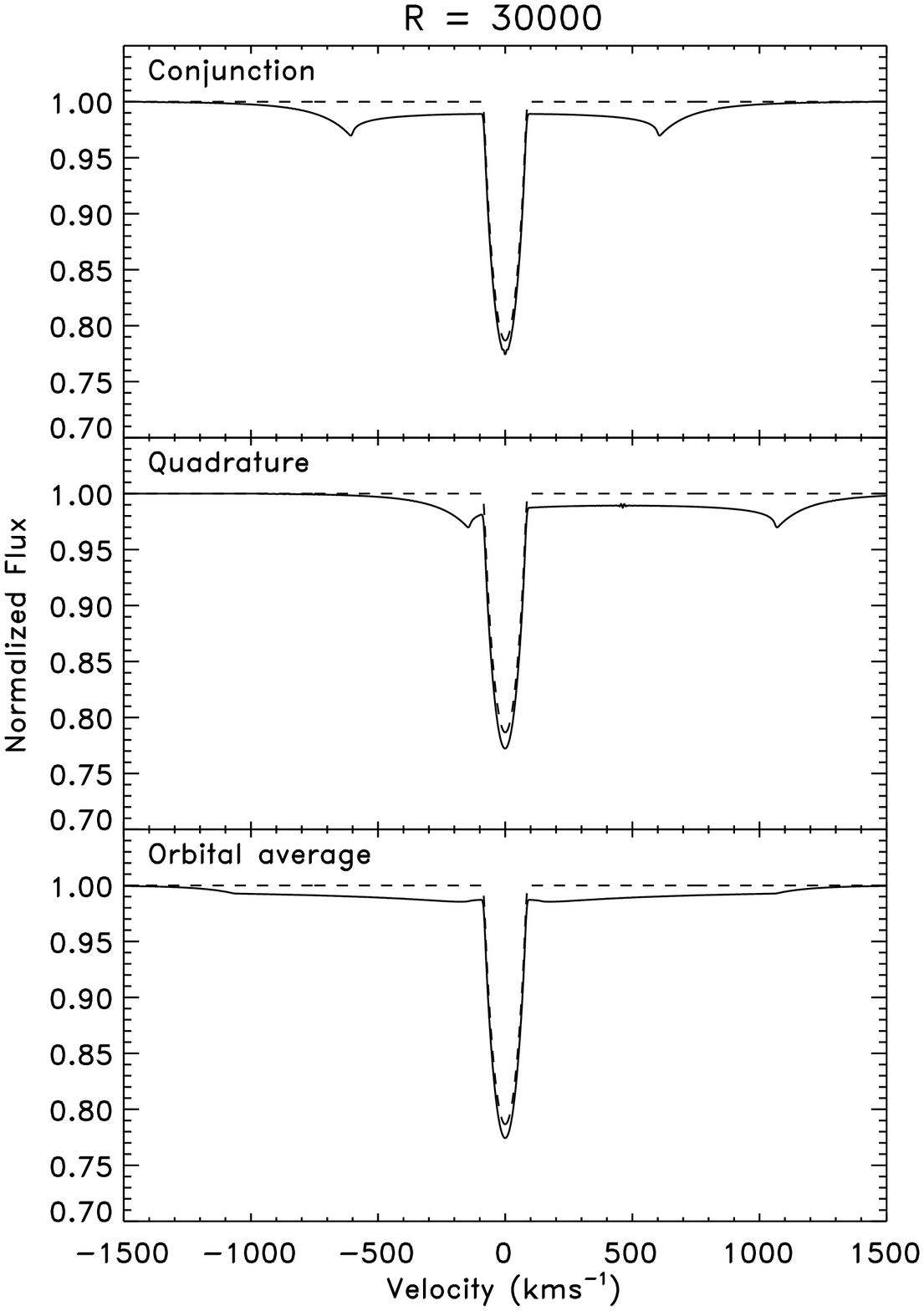}
\epsfig{width=2.4in,file=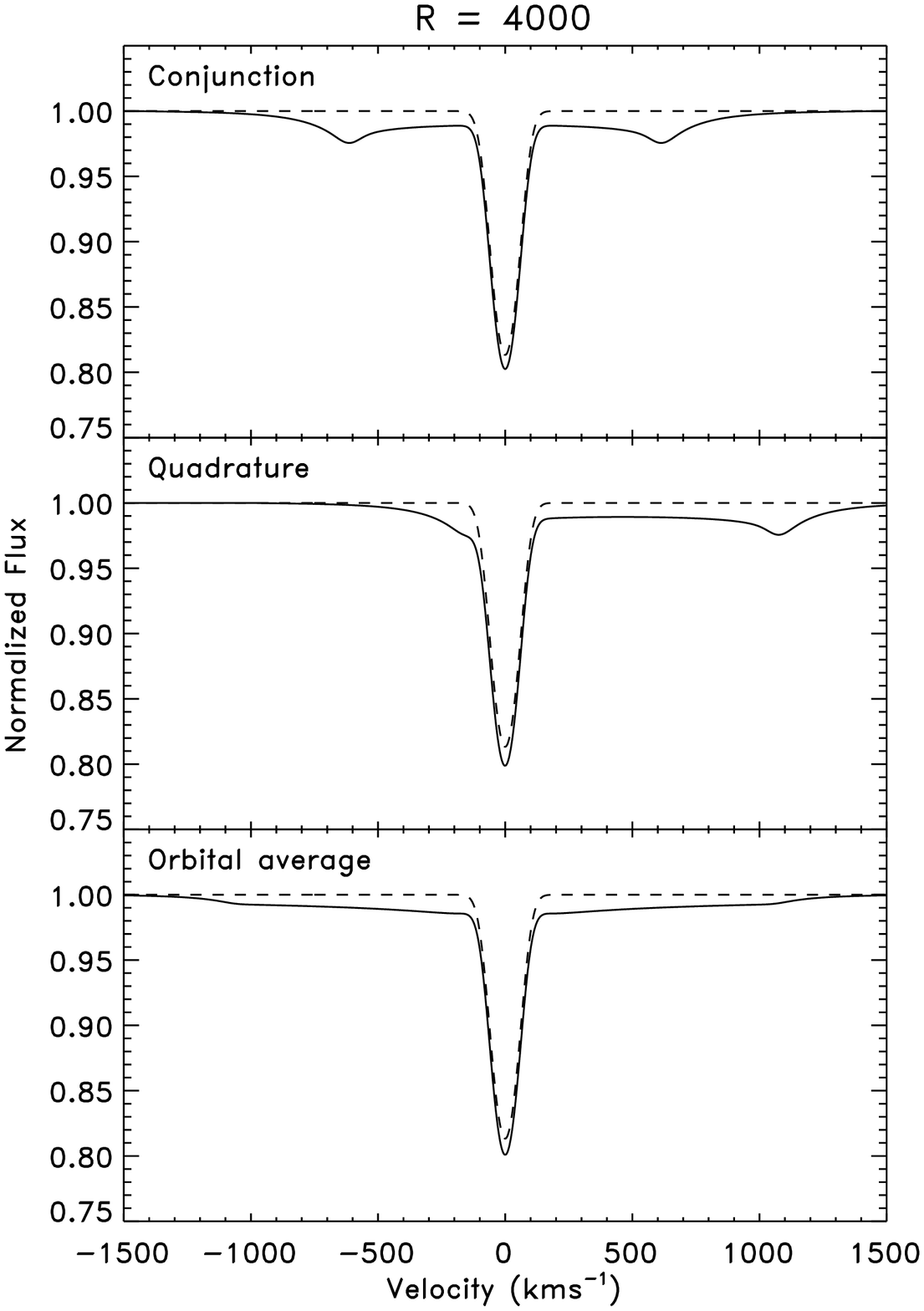}
\epsfig{width=2.4in,file=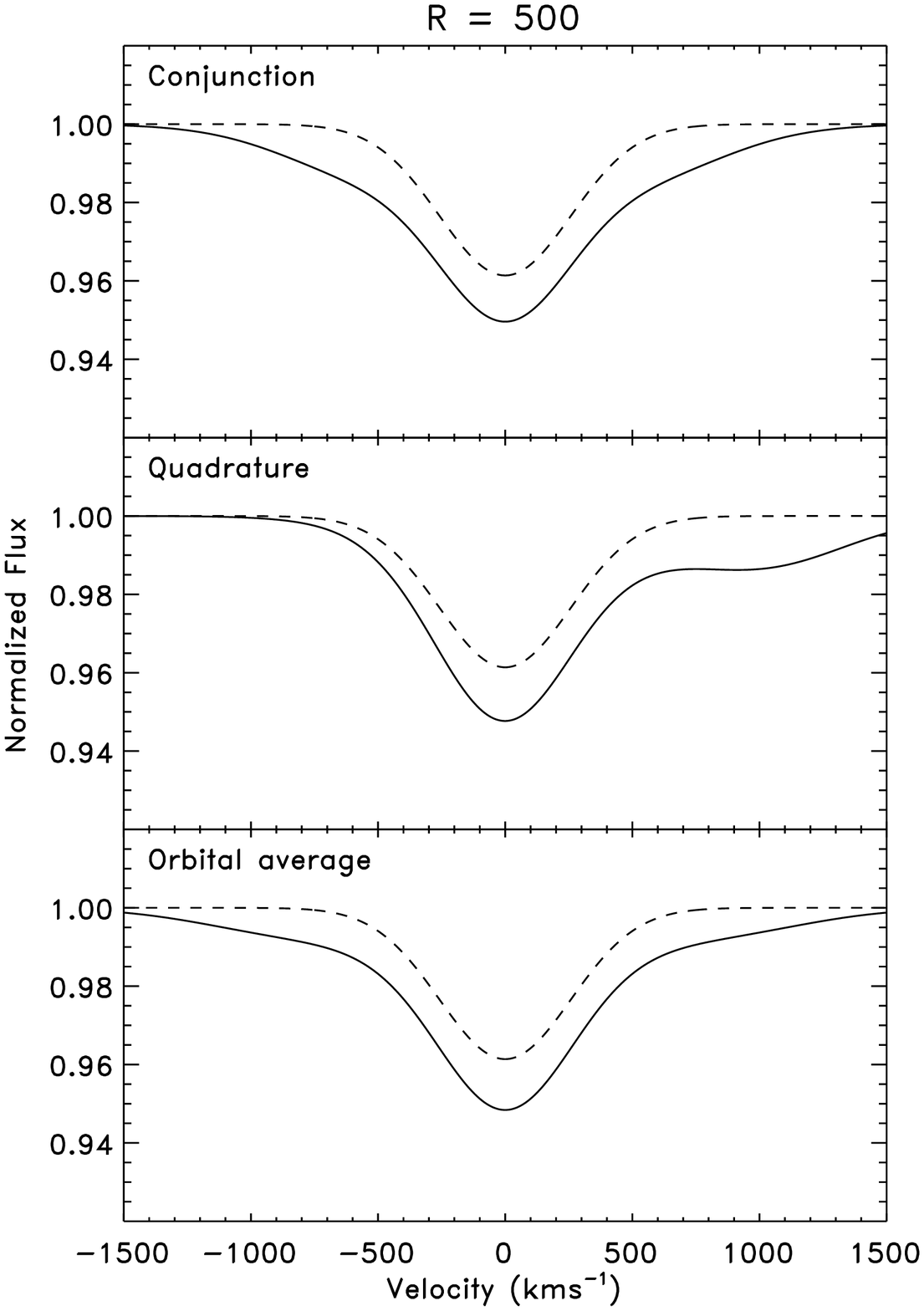}
\caption{Examples of composite line profiles calculated in the
companion star's rest frame for various resolutions.  We have used
parameters appropriate to \novamon, except that we have increased the
disk contribution to 50\,\%, so that star and disk contribute equally.
The solid line shows the profile derived if the disk has a line of
identical equivalent width to that of the companion.  The dashed line
assumes the disk is a featureless continuum.  Resolutions are chosen
to indicate cases corresponding to an arbitrarily high resolution
($R=30\,000$), a typical optical resolution ($R=4\,000$, e.g.\
\citealt{Marsh:1994a}), and a typical IR resolution ($R=500$, e.g.\
\citealt{Shahbaz:1999a}).}
\label{ProfileFig}
\end{figure*}


We also require a surface gravity.  We adopt the predicted value of
\citet{Gelino:2001a} of $\log g = 4.46\pm0.04$, based on the derived
system parameters.  It is consistent with the measured (spectroscopic)
value of $4.2\pm0.3$.  The small uncertainty in $\log g$ is dominated
by that in the mass ratio of \citet{Marsh:1994a} of $q =
0.067\pm0.010$.  If the inclination is larger than that derived by
\citet{Gelino:2001a}, for example if IR veiling is significant, then
the effect on $\log g$ is relatively small.  For $i=50^{\circ}$ and
$60^{\circ}$ it decreases to 4.40 and 4.35 respectively.

Finally for the companion we require a metallicity and rotational
broadening velocity.  For the metallicity we assume solar, since this
is the default assumption for most systems.  \citet{Gonzalez:2004a}
actually find slightly super-solar metallicity ($0.14\pm0.2$), but
this is consistent with solar within uncertainties.  For the
rotational broadening we adopt 83\,km\,s$^{-1}$ from
\citet{Marsh:1994a}.

To compute disk line profiles we additionally require orbital
parameters.  For this purpose we assume $P_{\rm orb}=7.78$\,hrs,
$q=0.067$, $M_1=11$\,M$_{\odot}$, and $i=40.75^{\circ}$
(\citealt{Gelino:2001a} and references therein).

Veiling measurements have been made in both optical and IR for
\novamon.  \citet{Marsh:1994a} measured disk contributions of
$6\pm3$\,\%\ at \Halpha\ and $17\pm3$\,\%\ at \Hbeta\ using a K3--K4V
template.  \citet{Gonzalez:2004a} found comparable values
($0\pm5$\,\%\ and $20\pm5$\,\%\ respectively).  \citet{Shahbaz:1999a}
obtained an upper limit on the disk contribution in the IR of
$27$\,\%\ ($2-\sigma$).  This is not a robust result, but it is all we
currently have.  Over and above the limitations discussed explicitely
later in this work, two major concerns are i) the data quality is not
high enough to be confident that the main CO features are securely
detected, and ii) cataclysmic variable star secondaries exhibit
anomalously weak CO features, possibly due to prior CNO processing
\citep{Harrison:2004a}.  The same mechanism may operate in BHXRT
companion stars (\citealt{Ergma:2001}; \citealt{Haswell:2002a}),
calling into question the usefulness of CO bands for veiling
measurements.

\section{The Simulations}

\subsection{Methodology}

In general, our analysis technique is to create model composite
spectra including a rotationally broadened model atmosphere spectrum
for the companion, and either a flat continuum or a model atmosphere
for the disk, convolved with a disk line profile.  Throughout we aim
to reproduce the main characteristics in order to obtain illustrative
results; we are not attempting precise, state-of-the-art modeling of
the quiescent spectra.  The synthetic spectrum can be expressed as
$(1-\alpha) F_{\rm star} + \alpha F_{\rm disk}$, where $\alpha$ is the
assumed true disk fraction (veiling), $F_{\rm star}$ is the input
stellar spectrum, and $F_{\rm disk}$ is either unity everywhere, or an
input disk spectrum.  Having constructed the model spectrum in this
way, we then fit it with $(1-\beta) F_{\rm template} + \beta$, where
$\beta$ represents the derived veiling and $F_{\rm template}$ is the
spectrum of the adopted template star.  Note that in fitting templates
to model data we always assume the disk is a flat continuum, following
observational practice.  The difference between $\alpha$ and $\beta$
then indicates how much the veiling is biased by choice of template,
or the presence of disk spectral features.  Our treatment is intended
to represent the ideal observational case, so we use noise-free
normalized spectra.  In analyzing actual observations further
uncertainties will be introduced by noise and difficulties in
normalizing the spectra.

We use two representative wavelength ranges.  For the optical we use
5920--6520\,\AA; this excludes the strong NaD lines as these might be
contaminated by interstellar absorption in observational data, and
also avoids the H$\alpha$ and He\,{\sc i} 5875\,\AA\ emission lines.
In the IR we use 2.19--2.31\,$\mu$m, to exclude Br$\gamma$, but
include the first CO bandhead (as used by \citealt{Shahbaz:1999a}).
We calculate models appropriate to three instrumental resolutions,
represented by Gaussians.  The reference case uses $R = \lambda /
\Delta\lambda = 30\,000$, sufficient to resolve the stellar rotational
broadening profile.  The second case is $R=4\,000$, representative of
optical studies, and based on the specific example of
\citet{Marsh:1994a}.  The third case is $R=500$, based on the IR
veiling determination for \novamon\ by \citet{Shahbaz:1999a}.

\subsection{Synthetic Line Profiles}

For convenience we calculate all model line profiles in the rest frame
of the companion star; in practice observers often shift spectra into
this frame and average, so this is appropriate.

We represent the stellar line profile with a rotational broadening
profile \citep{Gray:1992a} assuming linear limb-darkening with a
limb-darkening coefficient of 0.73 in the optical, appropriate around
6000--6500\,\AA\, and 0.31 in the IR \citep{AlNaimiy:1978a}.  This is
obviously a crude approximation to a distorted Roche-lobe filling
star, but is adequate for our purposes.  Likewise, linear
limb-darkening is imprecise, but commensurate with other
approximations made.

Disk profiles assume a disk extending from $R_{\rm in}$ to $R_{\rm
out}$. $R_{\rm in}$ is taken to be $10^4$ Schwarzschild radii, but in
practice has negligible effect on the line profiles.  We expect
$R_{\rm out}$ to lie between the circularization radius, the minimum
radius at which new material is likely to be added to the disk, and
the tidal truncation radius, the approximate maximum size.  The former
may be more likely for a quiescent disk, but the latter produces
narrower profiles, and hence is more likely to contaminate the stellar
spectrum.  For A\,0620--00, these assumptions yield $v_{\rm out} \sin
i = 600$\,km\,s$^{-1}$ for the tidally truncated case and $v_{\rm out}
\sin i = 740$\,km\,s$^{-1}$ for a disk extending out to the
circularization radius; we adopt the narrower tidally truncated
profile.  For simplicity we assume a disk with uniform temperature and
surface gravity and hence no variation in the emergent spectrum with
location.  This is a gross simplification but probably no worse an
approximation than the one we make by using stellar atmosphere spectra
in place of more realistic disk atmospheres.  The temperature
distribution in quiescent disks is in any case expected to be rather
flatter than in outbursting systems (\citealt{Lasota:2001a} and
references therein).  Note that since we are calculating line profiles
in the companion star's rest frame, the disk appears to move with
velocity semi-amplitude $K_1+K_2$ (where $K_1$ and $K_2$ are the
velocity semi-amplitudes of the compact object and companion star
respectively).

We show model composite line profiles in Fig.~\ref{ProfileFig},
assuming equal fluxes from star and disk, and lines of the same
equivalent width.  Examining the high resolution case first, it can be
seen that the much higher rotational broadening in the disk has the
effect of completely separating the two components, even though the
local emergent lines were assumed to be identical.  The effect is
amplified even more when the orbital average is considered, as this
further smooths the disk profile out.  This separation is well
preserved for a typical optical resolution ($R\sim4\,000$), and the
two components are sufficiently decoupled that the apparent depth of
the optical absorption line is little affected.  The disk line is
effectively broadened by so much that it appears almost as a smooth
continuum on the scale of the stellar features.  At even lower
resolutions the separation becomes less clear, however.  There are
obviously still two components, but they merge more and one might
expect a more significant contamination.  Nonetheless, if one could
isolate a single profile at high enough signal-to-noise, the stellar
component could be separated and measured reasonably reliably.  In
practice this is not what is done of course, and the broad disk
components would prove more difficult to separate when multiple
blended lines are present and when one must rely on statistical
measures, such as $\chi^2$ fitting, to estimate the veiling.

\subsection{Synthetic Spectra}

We next convolve the above line profiles with model atmosphere
spectra.  We construct these using atomic and molecular line data and
a grid of local thermodynamic equilibrium (LTE) ATLAS9 model
atmospheres obtained from the Kurucz
website\footnote{http://kurucz.harvard.edu}.  We interpolate within
these models for a given set of $T_{\rm eff}$, $\log g$, and $[Fe/H]$
and then compute model spectra using the LTE stellar synthesis code,
MOOG \citep{Sneden:1973a}.

\begin{figure}
\epsfig{angle=90,width=3.4in,file=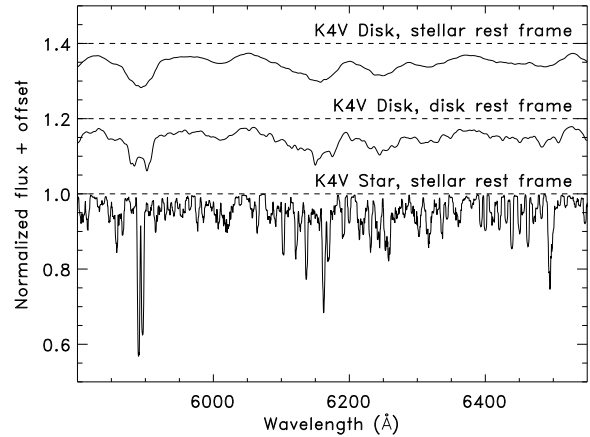}
\caption{Simulated spectra of a K4V star, representative of the
  companion to \novamon, and a disk with the same temperature and
  gravity.  This illustrates how severely the spectrum is modified by
  the disk's rotational broadening (in the disk rest frame), and
  orbital smearing (in the stellar rest frame).  Only the strongest
  features remain identifiable, and the distinctive double-troughed
  disk line profile is only visible in the disk rest frame.
  Successive vertical offsets of 0.2 units have been applied.}
\label{AtmosphereFig}
\end{figure}

Fig.~\ref{AtmosphereFig} shows examples of stellar and disk spectral
models calculated in this way for the same assumed temperature and
gravity, to illustrate the gross differences between the two spectra,
and reinforce the point made in the previous section.


\section{Systematic Errors in Veiling Estimates}

\subsection{Companion temperature}

A serious problem with the veiling estimates arises from the
uncertainty in the spectral type.  These estimates are extremely
sensitive to a mismatch between the template and the data, as is
illustrated for GS\,2000+25 by \citet{Harlaftis:1996a}, for example,
where disk contributions of 0--35\,\%\ are inferred for K1--M0
templates.  In theory, the best fitting template will indicate the
true spectral type, but the correct choice may not be obvious from
$\chi^2$ values alone, as is the case for the GS\,2000+25.

\begin{figure}
\epsfig{angle=90,width=3.4in,file=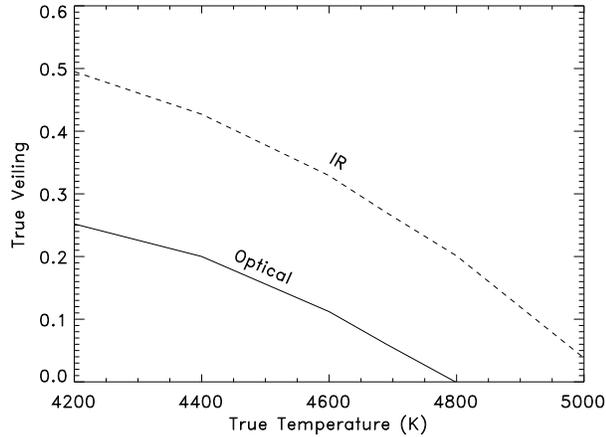}
\caption{True veiling, $\alpha$, required to reproduce observed values
if the template is a poor match in temperature to the star.
Calculations assumed an observed optical veiling of 6\,\%\ (solid
line), and an upper limit of 27\,\%\ in the IR (dashed line), and a
4690\,K template.}
\label{TempErrorFig}
\end{figure}

To further explore this problem we perform a series of simulations in
which we fit the $T_{\rm eff}=4690$\,K, $\log g = 4.46$ template plus
featureless disk to composite models containing stellar spectra of a
range of temperatures plus a featureless disk.  In each case, we
adjust the assumed (true) veiling, $\alpha$, until the derived
veiling, $\beta$, matches that observed.  Fig.~\ref{TempErrorFig}
shows the assumed veiling required for a range of stellar temperatures
roughly corresponding to the temperature uncertainty in A\,0620--00.
Large errors are clearly possible: if the true temperature is just
100\,K smaller than that of the template then the optical veiling is
underestimated by a factor of 2, and if it is 100\,K larger then no
veiling is required at all.  Indeed \citet{Gonzalez:2004a} derived
zero red veiling in A\,0620--00 for their preferred $T_{\rm
eff}=4900$\,K solution.  If the measured veiling is larger then the
fractional error is less significant, but will still usually exceed
the statistical error, and will probably always dominate the
uncertainty in veiling measurements.  Similar effects occur in the IR,
where the quoted upper limit of
\citet{Shahbaz:1999a} could be consistent with a true veiling of
up to 50\,\%\ for severe template mismatch.


In both the optical and IR, if the temperature of the template is
lower than that of the star, the veiling will be overestimated because
cooler templates generally have stronger lines (averaged over the
wavelength region being fitted).  In the case of A\,0620--00, there
are indications that that may be the case, as discussed above, so
existing veiling estimates are probably useful as upper limits.  In
most systems, however, we do not have such a precise temperature and
cannot make such an assumption.

\subsection{Companion gravity}

As noted above, the uncertainty in gravity for A\,0620--00 is not
large.  Nonetheless, the CO bands in particular are expected to be
sensitive to gravity, so even a small uncertainty might be
significant.  Furthermore, the gravity of A\,0620--00, and most BHXRT
companions, is lower than that of a main sequence star, and hence
lower than typical templates.  For example \citet{Gray:1992a} gives
$\log g=4.60$ for a K4V star.

As above, we take a $T_{\rm eff}=4690$\,K, $\log g = 4.46$ template
plus featureless disk model and use it to fit synthetic optical
spectra comprising a star of the same temperature but different
gravities, plus a 6\,\%\ featureless disk contribution.  An
uncertainty in $\log g$ of 0.04 has negligible effect, and allowing a
range of $\pm0.1$ still yields derived veilings of $6\pm2$\,\%, within
the statistical uncertainties of \citet{Marsh:1994a}.  The most
extreme case is when we use a main-sequence $\log g = 4.60$ template
and fit to a $\log g = 4.35$ spectrum.  Even then a derived veiling of
6\,\%\ corresponds to a true veiling of 2\,\%, close to the edge of
the quoted error range.  Therefore uncertainties in gravity, and
mismatch with a main-sequence template, are not the dominant source of
error in the optical.

In spite of the gravity sensitivity of the CO bandheads, there is also
negligible effect in the IR.  Using a $\log g=4.46$ template, we a
mismatch in gravity only raises the upper limit from 27\,\%\ to
29\,\%\ for gravities of $\log g=4.46\pm0.10$.  Fitting a $\log
g=4.60$ template to a model spectrum with gravity of $\log g=4.35$
only raises the limit to 30\,\%. Note that the bias is in the opposite
sense to that found in the optical.

\subsection{Companion metallicity}

Another difficulty is that the line strengths will also be a function
of the metallicity.  A mismatch between the metallicities of the
template and companion could also introduce an error in the veiling
measurement; a low metallicity companion will have weaker lines, and
hence appear more heavily veiled than it actually is.

The metallicity of the secondaries of BHXRTs is not well constrained
by observations.  \citet{Kotoneva:2002a} examine metallicities of
nearby K dwarfs and find a spread in ${\rm [Fe/H]}$ of approximately
$-1.0$ to $+0.2$.  The distribution peaks at -0.2.  For comparison,
\citet{Gonzalez:2004a} derive ${\rm [Fe/H]}=+0.14\pm0.20$ for
A\,0620--00, at the high end of the range.  Given this range, unless
the template star is deliberately selected by metallicity, a mismatch
is not just possible, but likely.  The template used by
\citet{Marsh:1994a}, HD\,16160, has ${\rm [Fe/H]}=-0.03\pm0.14$
\citep{Heiter:2003a}, so the mismatch is small, but not negligible
($0.17\pm0.24$).  In the case of the IR measurement of
\citet{Shahbaz:1999a}, however, the preferred template (HD\,42606)
does not have a published metallicity, so a larger mismatch is
possible.

To test the sensitivity to metallicity we adopt a model spectrum of
$T_{\rm eff}=4690$\,K, $\log g=4.46$, solar metallicity, and a
variable disk contribution.  The template we fit to the model is
assumed to have the same temperature and gravity, but metallicities of
-0.3, -0.6, and -0.9, in accord with the range found by
\citet{Kotoneva:2002a}.  With these combinations we find that to
reproduce an observed optical veiling of 6\,\%\ requires true veilings
of 20\,\%, 32\,\%, and 46\,\%.  In the IR the sensitivity to
metallicity is also large; an upper limit on the veiling of 27\,\%\
permits true veilings of up to 41\,\%, 54\,\%, and 67\,\%\ for
template metallicities of -0.3, -0.6, and -0.9 respectively.  Thus if
the template has lower metallicity than the target, then the veiling
could have been substantially underestimated.

Returning to \novamon, \citet{Gonzalez:2004a} obtain a metallicity of
$[Fe/H]=0.14\pm0.20$, whereas the template used by \citet{Marsh:1994a}
has a metallicity of $[Fe/H] = -0.03\pm0.14$.  On the one hand the
difference between them is not statistically significant, but on the
other it could be large enough to bias the veiling measurement
substantially, by a factor of two or more.  Mismatch between template
and target metallicities can thus be a major contributor to the
uncertainty in veiling measurements.

\subsection{Disk spectral features}

\begin{figure}
\epsfig{angle=90,width=3.4in,file=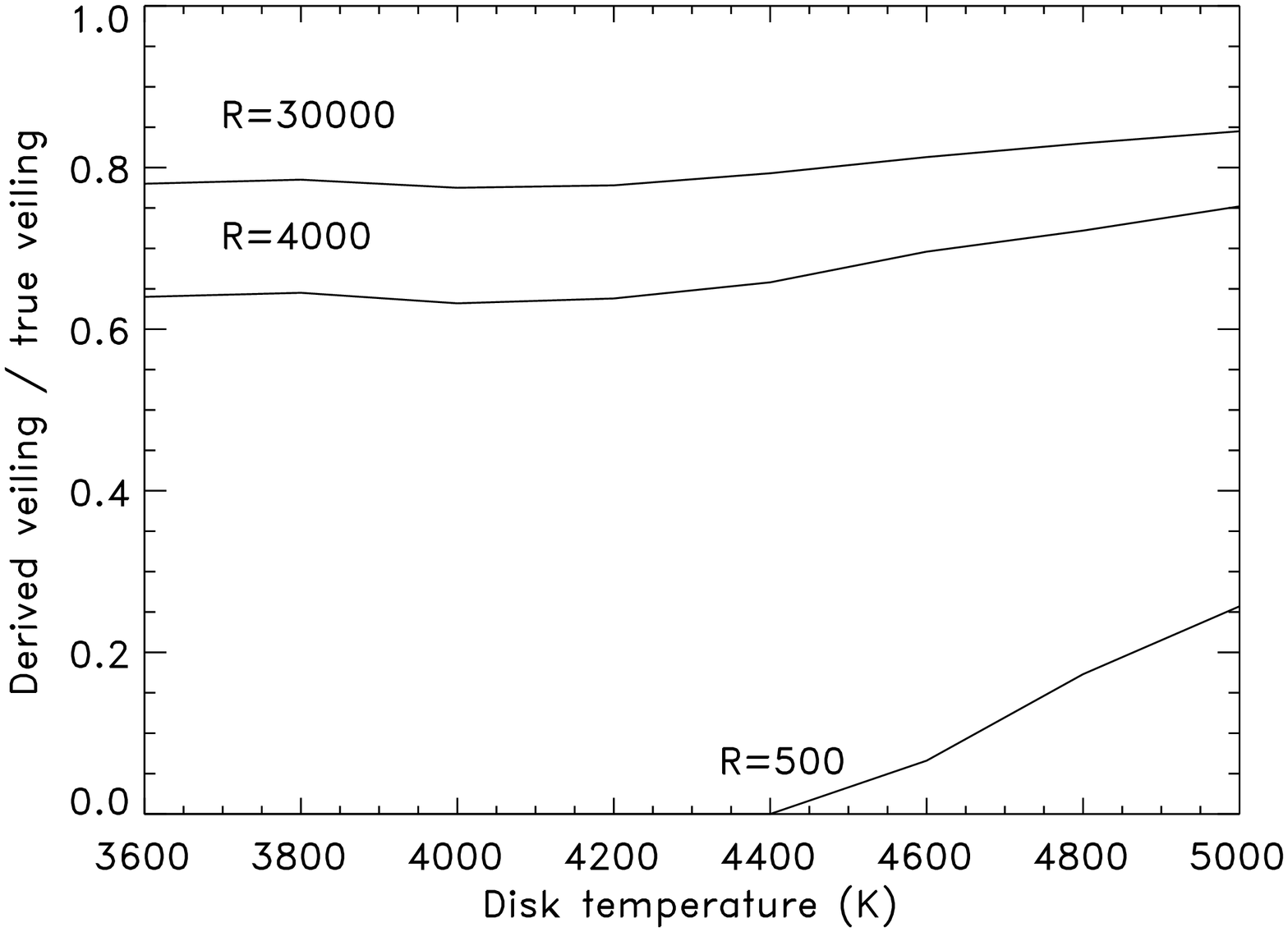}
\epsfig{angle=90,width=3.4in,file=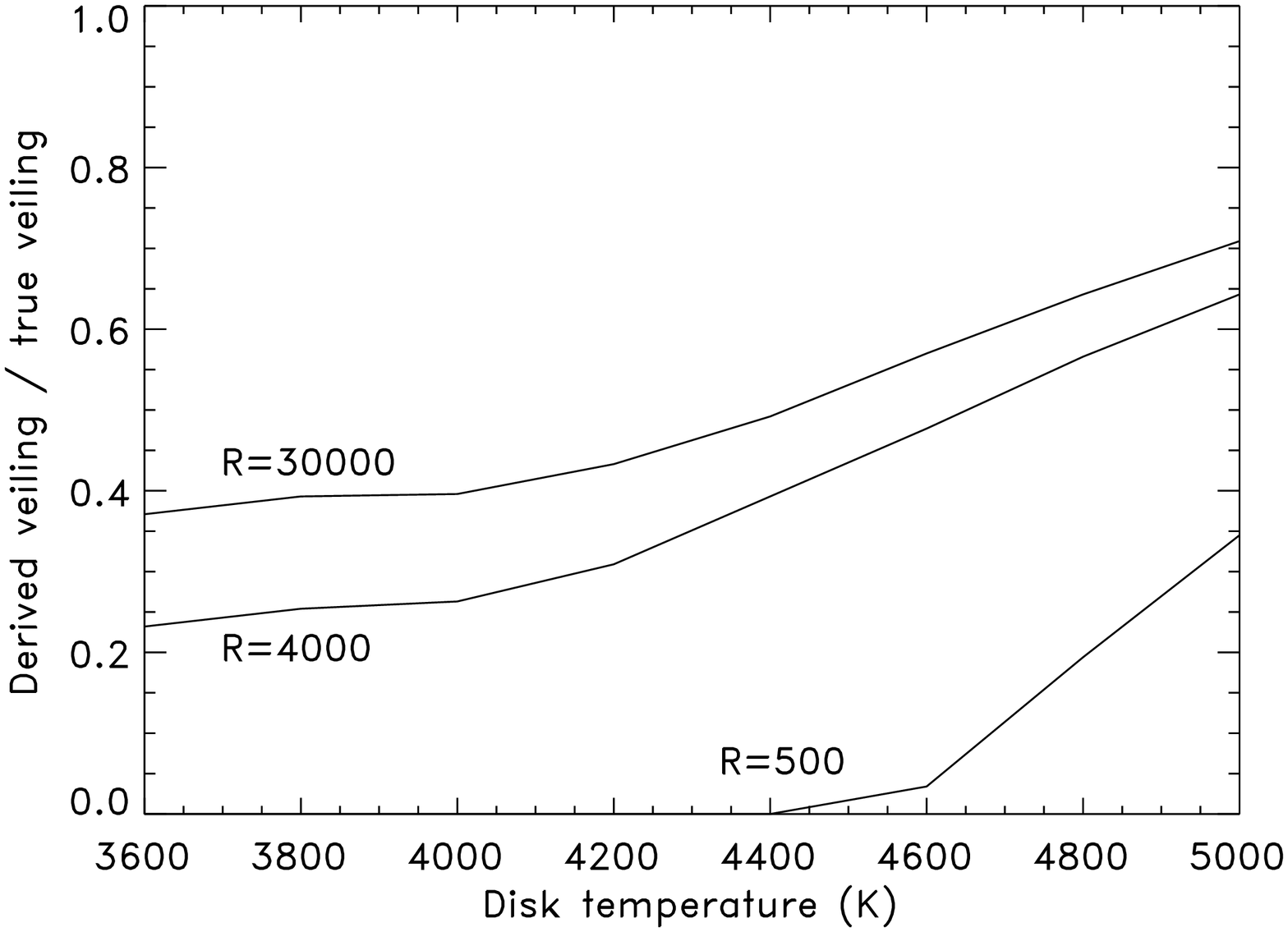}
\caption{Effect on veiling estimates if the disk has absorption
  lines.  The upper panel uses a red spectrum (5920--6520\,\AA), and
  the lower panel a section of the $K$ band (2.19--2.31\,$\mu$m).  
  This is for an ideal case where the template spectrum is the
  same as was used in the simulation.  Even in this case, significant
  errors occur, especially at low resolution, or when broad bands such
  as CO dominate as is the case in the $K$ band.}
\label{VeilFig}
\end{figure}

We finally proceed to test the robustness of veiling estimates if the
disk spectrum is not featureless.  We consider an ideal case, where
the spectra are averaged over a full binary orbit in the rest frame of
the companion.  We use the modeled companion star spectrum ($T_{\rm
eff}=4690$\,K, $\log g=4.46$) as the template for the stellar
component so that no bias is introduced by template mismatch.  Values
of $\beta/\alpha$ are shown in Fig.~\ref{VeilFig} for various assumed
disk temperatures and resolutions; we have used $\alpha=0.1$ for these
tests.


We consider first the optical case.  For moderately high resolutions
($R=4,000$ and $R=30,000$), the error in the veiling is modest, and
would usually be within the statistical uncertainties in the veiling
measurement (which are typically large).  The error is larger for a
cool disk because the spectral features tend to be deeper.  At low
resolutions, the problem becomes much more severe.  In this case, as
can be seen in Fig.~\ref{ProfileFig}, the disk and stellar line
profiles are no longer well resolved, and it becomes hard for the
fitting algorithm to distinguish between them, even with a template
perfectly matched to the star.  For low disk temperatures, the
stronger lines actually mean that adding a disk contribution makes the
overall lines appear stronger rather than weaker, and one would derive
a negative veiling (clipped here to zero).

In the IR the situation is less favorable, at least if the primary
diagnostics are CO bands.  Data are typically obtained at lower
resolution than in the optical (as in \citealt{Shahbaz:1999a}), and
even if higher resolutions are used, the intrinsic width of the CO
bands limits the discrimination possible between disk and stellar
profiles.  The low resolution case is thus comparable to the same
resolution in the optical, but there is less benefit to high
resolutions, unless the study focuses only on metallic features.  As
noted earlier, contamination of the CO bands by the disk might be {\em
expected} in quiescent BHXRTs, as CO emission is seen in the quiescent
cataclysmic variable WZ~Sge \citep{Howell:2004a}, although it is not
clear that the cases are comparable.  If contamination is by emission
features then the veiling will tend to be overestimated.

We thus conclude that provided veiling studies are done at
intermediate resolution or better, the bias on the veiling is not the
dominant source of error in the optical, even if the disk has strong
absorption lines.  At low resolutions, such as commonly used in the
IR, the measurement could be severely compromised.  This will also
apply (irrespective of resolution) if the features used are not
narrow, e.g.\ CO or TiO molecular bands.

\section{Effect on the Broad-band Spectral Energy Distribution}

It is usually argued (e.g.\ \citealt{Gelino:2001a}) that since the
veiling is measured to be lower in the red than in the blue, then the
IR veiling must be even lower.  The implicit assumption being made
here, however, is that the veiling varies monotonically with
wavelength.  There is no requirement that this be the case; multiple
components may contribute to the disk spectral energy distribution.
The optical veiling (and UV spectrum) may well be dominated by a
stream impact point or overflow \citep{McClintock:2003a}, and appears
to be relatively blue.  The stream-impact point certainly does appear
as a bright source of emission line light in Doppler tomograms
\citep{Marsh:1994a}.  In contrast, it is observed that rapid flaring
exhibits a rather redder spectrum, comparable to the color of the
companion star (\citealt{Shahbaz:2003a}); this could represent the
optically thin disk component.  Optically thick disk material could
also be present.  This material is predicted to be very cool
($\sim$3000\,K) and will not contribute much flux at optical
wavelengths.  Given the large projected surface area of the disk
(several times that of the companion), an extended but cool component
might, however, produce a significant contribution in the IR, even if
it is undetectable in the optical.  One can then imagine a disk with
two components, a hot component dominating optical and UV wavelengths
and a cool component dominating infrared wavelengths.  Such a picture
is consistent with theoretical expectations, but has the unhappy
consequence that the total veiling may well be minimized in the red,
precisely where most measurements are made!

We can attempt to quantify the possible contribution of the cool disk
material to the IR by considering optically thick disks with a range
of temperatures and projected areas.  As above, we assume a uniform
disk temperature and parameterize the area as the covering factor of a
tidally truncated disk.  For \novamon\ such a maximal disk has a
projected area $\sim7$ times that of the companion star, assuming
$i=40\fdg75$ \citep{Gelino:2001a}, whereas a disk only extending to
the circularization radius still subtends an area $\sim3$ times that
of the companion.  We can then estimate the maximum covering factor
which is consistent with the observed optical veiling as a function of
temperature.  For this purpose we now use the most robust estimate, a
red veiling of $0\pm5$\,\%\ \citep{Gonzalez:2004a}.  We estimate the
relative contributions from the companion and disk as a function of
temperature by using surface brightness from NextGen model atmosphere
calculations \citep{Hauschildt:1999a}.  This of course explicitly
assumes the disk is optically thick, so these constraints will not
apply if the disk is optically thin.


\begin{figure}
\epsfig{angle=90,width=3.4in,file=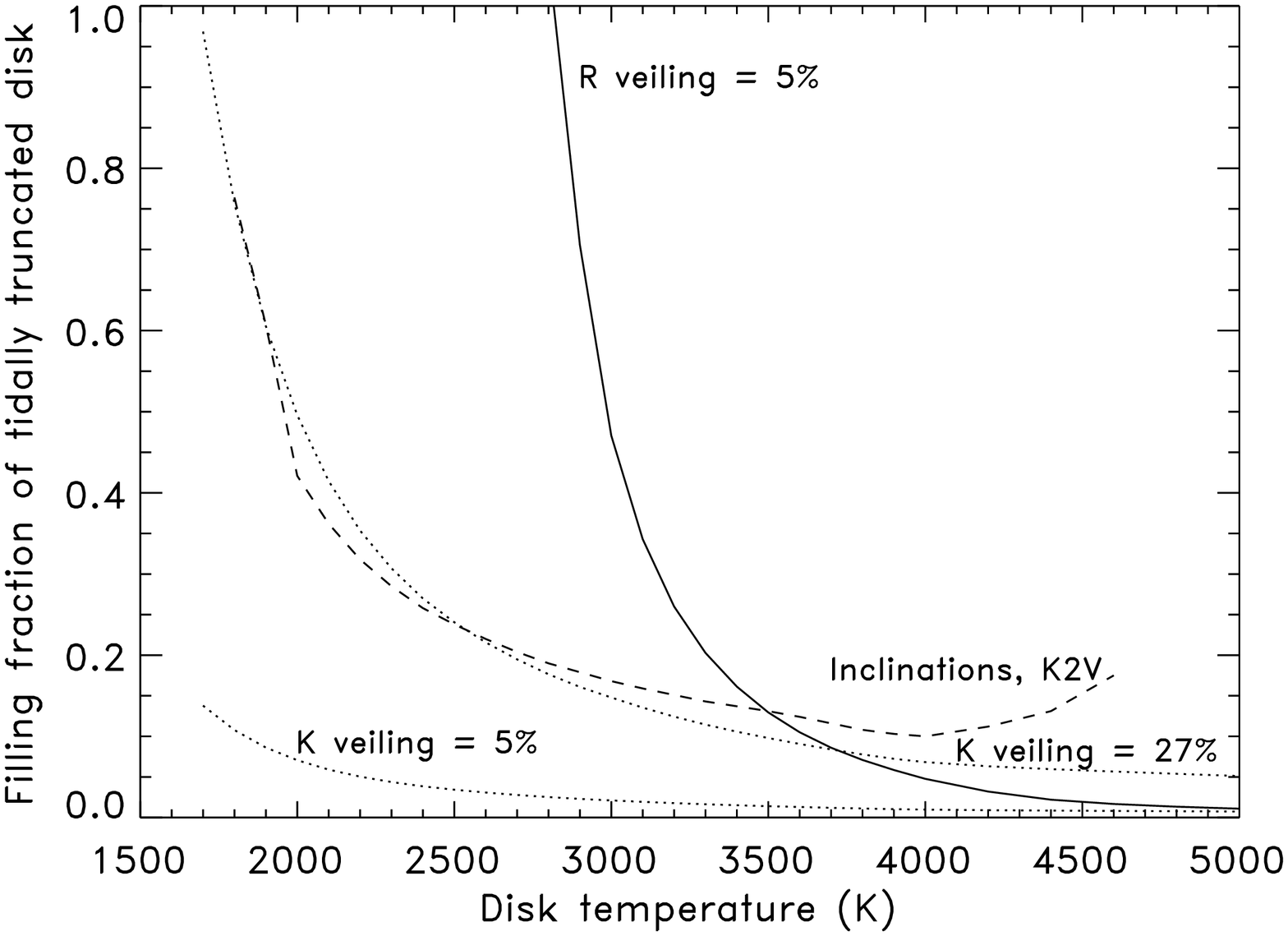}
\caption{Upper limits on the temperature and covering fraction
  (relative to a disk extending to the tidal truncation radius) of an
  optically thick disk in \novamon.  The solid line indicates
  constraints due to the observing veiling in the red
  \citep{Gonzalez:2004a}.  Dotted lines are based on the veiling at
  $K$ \citep{Shahbaz:1999a} and dashed lines are based on the
  consistency of inclination determinations across $J$, $H$, and $K$
  \citep{Gelino:2001a}.  The veiling limits assume a K4V spectral type
  as this places stricter limits than earlier types.  Conversely, the
  inclination constraint is based on K2V, as this is stricter than
  later types.  Veiling constraints correspond to $2\,\sigma$ limits
  ($\sim90$\,\% confidence).  The inclination constraint uses
  $1\,\sigma$ limits, but is effectively based on a joint confidence
  region in the $J$ and $K$ (with their errors in opposite
  directions), so is also close to 90\,\%\ confidence.}
\label{CoverFig}
\end{figure}


We find that disks of a temperature comparable or greater than the
companion are highly constrained by the optical veiling if this is
reliable (see Fig.~\ref{CoverFig}); this constraint would become even
tighter if one allows for the presence of a blue veiling component as
well.  What this means is that the quiescent disk must either be much
cooler than the companion ($\la3500$\,K) or mostly optically thin
(with perhaps small clumps of optically thick material).  Note that
since the circularization radius effectively limits the minimum size
of the disk, one cannot explain this away entirely with a small disk;
the minimum covering factor possible in this way is about 0.4.

As the disk temperature is allowed to drop, however, the constraint
relaxes, and for theoretically motivated temperatures
($\sim$3\,000\,K; e.g.\ \citealt{Menou:2002a}) a more respectable
covering fraction is possible.  Such a disk can actually contribute
strongly in the $K$ band while having negligible effect in the
optical.  It would of course veil the features of the companion star,
so $K$ band veiling estimates could still constrain its presence, but
as we have seen the existing IR estimate \citep{Shahbaz:1999a} is
potentially compromised by even a small mismatch with the template
temperature or metallicity, or by disk spectral features.  The upper
limit derived, 27\,\%\ at 2-$\sigma$, however, does not strongly
constrain a cool disk (Fig.~\ref{CoverFig}), especially when we
consider that the upper limit may not be very robust.

An additional constraint comes in the form of the consistency of
inclination determinations from the simultaneous $J$, $H$, and $K$
lightcurves (\citealt{Gelino:2001a}; Gelino 2005, private
communication).  These authors find inclinations derived from
individual bands are all consistent with the combined fits to within
uncertainties ($\pm 1.5^{\circ}$).  If veiling were present with
colors different to those of the star, then different bands would
experience different veilings and hence discrepant inclinations would
be derived.  We show in Fig.~\ref{CoverFig} the constraint this places
on the disk covering factor with a K2V companion star.  Assuming K4V
results in slightly higher limits.  Unfortunately even these
observations only limit the covering factor, and hence veiling, to a
similar precision as claimed from spectroscopic veiling.

A final possible constraint would come from the observed spectral
energy distribution (SED).  Indeed one of the arguments made by
\citet{Gelino:2001a} for negligible IR contamination was that the
photometric SED was consistent with the spectroscopic spectral type
constraints.  This is less robust than the inclination constraint,
however, as it is very sensitive to the assumed spectral
classification of the companion star, and hence results in weaker
limits than the inclination constraint does.  These have not been shown.

We conclude that an extensive, cool, optically thick disk component in
\novamon\ cannot confidently be rejected based on any existing
observational evidence.  There is no reason such a disk should not be
present, and it could have as large a contribution in the $K$ band as
veiling measurements allow, 25\,\%, or even more.  We know of no
evidence to support the assertion that the $K$ band contamination is
as low as a few percent.

\section{Discussion}
\label{DiscussionSection}

A cool, optically thick disk component might be expected in quiescent
BHXRTs as it has long been argued that an entirely optically thin disk
is inconsistent with the DIM (see discussions by
\citealt{Cannizzo:1993a} and \citealt{Menou:2002a}).  The argument is
made that an optically thin disk requires extremely low surface
densities, for example \citet{Lin:1988a} estimated $\Sigma \la
1$\,g\,cm$^{-2}$.  In contrast, the DIM operates at critical surface
densities of order 100\,g\,cm$^{-2}$ (e.g.\ \citealt{Hameury:1998a};
\citealt{Lasota:2001a}).  Even if one believes that the DIM is not
correct, \citet{Menou:2000a} estimate that during a typical outburst
of \novamon\ a few $\times10^{24}$\,g are accreted, corresponding to
an {\em average} surface density pre-outburst of at least
20\,g\,cm$^{-2}$.

While it would appear that most of the quiescent disk mass must then
be in the form of optically thick material, there is an inconsistency
here.  \citet{Lin:1988a} were considering relatively high temperature
material (by quiescent standards) at temperatures above 6000\,K,
whereas disk models are now working with temperatures that drop to
2000--3000\,K in quiescence.  At these temperatures opacities are
expected to be lower (e.g.\ \citealt{Ferguson:2005a}), so higher
surface densities could remain optically thin under these conditions,
and this issue requires further examination.

As a crude test we considered the expected optical depth, based on
Rosseland mean opacities taken from \citet{Ferguson:2005a} of a range
of uniform slabs of temperature 2000--3000\,K, surface density
20--100\,g\,cm$^{-2}$, and thickness 1--4$\times10^{9}$\,cm,
corresponding approximately to $H/R$ values of 0.01--0.02 for the
outer disk of a typical quiescent BHXRT such as A\,0620--00.  At low
temperatures ($T\sim2000$\,K) this essentially depends only on surface
density, with optical depths 0.2--1.2 likely.  At higher temperatures
($T\sim3000$\,K) volume density becomes important, and a larger range
of optical depths from 0.1--3.8 are found for the parameters described
above.  We conclude that based on current knowledge of such disks
optically thick regions might or might not be present; they do not
appear to be inescapable as has sometimes been claimed, but should
still be considered a possibility, and observational constraints such
as discussed here are needed to rule it out.

As has long been appreciated, an optically thick disk is not
inconsistent with the presence of disk emission lines, as these likely
originate from a chromosphere above the disk; indeed this chromosphere
could account for most or all of the optical emission.  Alternatively
it may be that there are optically thin patches in the disk in between
optically thick clumps (e.g.\ \citealt{Vrielmann:2002a}).  Such a
picture would be appealing to explain the mirror eclipses seen in
quiescent dwarf novae, which imply that at least 20\,\%\ of the disk
(by radius) must be optically thin in the continuum
(\citealt{Littlefair:2001a}; see also \citealt{Froning:1999a}).  Note
that if the disk mass is clumpy then this will further increase the
likelihood of clumps being optically thick.

Any optically thick material must be cool.  For it to be stable on the
cool branch of the disk instability requires $T_{\rm eff} \la
6000$\,K.  Unless it consists of a few very dispersed clumps, or is
indeed optically thin, the optical veiling further constrains the
temperature to $T_{\rm eff} \la 3500$\,K.  This is consistent with
typical theoretical temperatures $T_{\rm eff} \sim 3000$\,K (e.g.\
\citealt{Menou:2002a}).  In this regime, as we have seen, the
observational constraints are extremely weak, and an extensive
optically thick component contributing a significant fraction of the
light in the $K$ band is possible.

It would therefore be an interesting constraint if the disk
contribution at $K$ is only a few percent.  As shown in
Fig.~\ref{CoverFig}, this would require the area of the optically
thick fraction of the disk to be small, implying most of the disk is
optically thin.  A significant optically thick covering factor (more
than 10\,\%) is only consistent with 5\,\%\ contamination for $T_{\rm
eff} \la 1800$\,K.  It is unlikely that quiescent disk material can be
this cool, as there will always be some heating by the companion star,
the stream impact, and the central X-ray source.

One argument against significant IR contamination is that optical
ellipsoidal modulations are distorted and variable, but the IR
modulations are much cleaner.  This argument is not conclusive,
however.  It is likely that the optical veiling originates from the
stream impact and/or overflow.  This material is highly
non-axisymmetric, and variable.  The optically thick material,
however, is likely to be in an axisymmetric and probably non-variable
disk, and so is likely to only contribute a steady source of light
that does not distort the ellipsoidal variations.

\section{Conclusions}

We have examined the observational constraints upon the presence of a
cool, optically thick component to the accretion disks of quiescent
BHXRTs.  Such a component is theoretically plausible, but has not yet
been detected.  We have shown that constraints from veiling of the
stellar photospheric spectrum are not very robust, as the veiling
derived is very sensitive to a mismatch between temperatures and/or
metallicities of the companion star and template, and to a lesser
extent to a surface gravity mismatch.  If the disk spectrum exhibits
similar spectral features to the star, for example CO bands in
emission or absorption, then this can further bias veiling estimates.

The current situation in the prototypical system A\,0620--00 is that
reasonably robust optical veiling measurements now exist, in the sense
that the veiling, temperature, gravity, and metallicity are all
determined simultaneously by high resolution spectroscopy
\citep{Gonzalez:2004a}, but that the IR remains poorly constrained.
The red veiling tightly constrains the temperature and or covering
factor of a warm disk, requiring either $T_{\rm eff}<3500$\,K, or a
very low covering factor ($\la10$\,\%) of optically thick material.
Cooler disks, as expected theoretically, are unconstrained by the
optical measurement.  The IR veiling measurements made so far are
sensitive to the uncertainties summarized above so are not robust.  A
more robust approach yielding similar constraints to the IR veiling is
the relative amplitudes of ellipsoidal variations
\citep{Gelino:2001a}, but these still allow contamination up to $\sim
25$\,\%.

Thus we conclude that no current measurements constrain the $K$ band
contribution from the accretion disk to better than about 25\,\%\ in
A\,0620--00, and consequently that significant dilution of IR
ellipsoidal modulations remains possible.  In particular, the
assumption that the IR contamination is only a few percent is not yet
supported by compelling evidence.  This issue has important
implications for black hole mass determinations.
\citet{Gelino:2001a}, for example commented that increasing the $K$
band disk veiling to 50\,\%\ would increase the derived inclination to
$60^{\circ}$.  The derived black hole mass would then drop from
11\,M$_{\odot}$ to below 5\,M$_{\odot}$!  This is an extreme case,
clearly a milder and more plausible IR veiling could still have
substantial impact on the derived black hole mass.

Finally we note that if the $K$ band disk contamination can be
observationally constrained to be only a few percent then this will
place very tight constraints indeed upon the temperature and/or
covering factor of optically thick material, providing a test of
quiescent disk models.

\acknowledgments

RIH is grateful to Dawn Gelino for helpful criticism of this
manuscript, to Rick Hessman for discussions of patchy accretion disks
in quiescent objects, to Tariq Shahbaz on optimal subtraction
techniques.  RIH acknowledges support for most of this work from NASA
through Hubble Fellowship grant \#HF-01150.01-A awarded by the Space
Telescope Science Institute (STScI), which is operated by the
Association of Universities for Research in Astronomy, Inc., for NASA,
under contract NAS 5-26555.

\clearpage




\end{document}